\documentclass[usenatbib,usedcolum,usegraphicx]{mn2e}

\title[
          Measurements of Streaming Motions of the Galactic Bar
      ]{
          Measurements of streaming motions of the Galactic Bar
                        with Red Clump Giants
 }

\author[               T. ~Sumi, L. ~Eyer and P. R.~Wo\'{z}niak
       ]
       {             T. ~Sumi,$^1$ L. ~Eyer$^2$ and P. R.~Wo\'{z}niak$^3$\\
    $^{1, 2}$Princeton University Observatory, Princeton, NJ 08544-1001, USA;
    e-mail: sumi@astro.princeton.edu, leyer@astro.princeton.edu\\
    $^3$Los Alamos National Laboratory, MS-D436, Los Alamos, NM 87545;
    e-mail: wozniak@lanl.gov\\
}
\date{Accepted 
      Received
      in original form }

\pagerange{\pageref{firstpage}--\pageref{lastpage}}

\begin{document}
\maketitle
\label{firstpage}

\begin{abstract}
We report a measurement of the streaming motion of the stars in the
Galactic bar with the Red Clump Giants (RCGs) using the data of the
Optical Gravitational Lensing Experiment II (OGLE-II).  We measure
the proper motion of 46,961 stars and divide RCGs into bright and
faint sub-samples which on average will be closer to the near and
far side of the bar, respectively.  We find that the far-side RCGs
(4,979 stars) have a proper motion of $\Delta <\mu> \sim 1.5 \pm
0.11$ mas\,yr$^{-1}$ toward the negative $ l $ relative to the
near-side RCGs (3,610 stars).  This result can be explained by stars
in the bar rotating around the Galactic center in the same direction
as the Sun with $v_b \sim 100$ km\,s$^{-1}$.  In the Disc Star (DS)
and  Red Giant (RG) samples, we do not find significant
difference between bright and faint sub-samples. For those samples
$\Delta <\mu> \sim 0.3 \pm 0.14$ mas\,yr$^{-1}$ and $\sim 0.03 \pm
0.14$ mas\,yr$^{-1}$, respectively.  It is likely that the average
proper motion of RG stars is the same as that of the Galactic
center.  The proper motion of DSs with respect to RGs is $\sim 3.3$
mas\,yr$^{-1}$ toward positive $ l $. This value is consistent with
the expectations for a flat rotation curve and Solar motion with
respect to local standard of rest.  RGs have proper motion 
approzimately equal to the average of bright and faint RCGs, which 
implies that they are on average near the center of the bar.  
This pilot project demonstrates that OGLE-II data may be
used to study streaming motions of stars in the Galactic bar.  We
intend to extend this work to all 49 OGLE-II fields in the Galactic
bulge region.
\end{abstract}

\begin{keywords}
%
Galaxy:bulge -- Galaxy:center -- Galaxy:kinematics and dynamics--Galaxy:structure
\end{keywords}

\section{Introduction}

Several groups have carried out gravitational microlensing
observations toward dense stellar fields, such as the Magellanic
clouds, the Galactic center and disc.  Until now, hundreds of events
have been found (EROS: \citealt{aub93}; OGLE: \citealt{uda93,uda94,uda00};
\citealt{woz01}; MACHO: \citealt{alc97,alc00a,alc00b}; MOA: 
\citealt{bon01,bon02}),
and thousands are expected in the upcoming years by MOA
\footnote{{\tt http://www.phys.canterbury.ac.nz/\~{}physib/alert/alert.html}},
OGLE-III
\footnote{ {\tt http://www.astrouw.edu.pl/\~{}ogle/ogle3/ews/ews.html}}
and other collaborations.

It is well known that the gravitational microlensing survey data is
well suited for numerous other scientific projects (see
\citealt{pac96}; \citealt{gou96}).  The studies of the Galactic structure
certainly benefit from this type of data.  The microlensing
optical depth probes the mass density of compact objects along the
line of sight and the event time-scale distribution is related to
the mass function and kinematics of the lensing objects.  Observed high
optical depth may be explained by the presence of the bar
(\citealt{uda94}; \citealt{alc97,alc00a}; \citealt{sum02}).  There is
substantial evidence that the Galaxy has a bar at its center
(\citealt{vau64}; \citealt{bli91}; \citealt{sta94,sta97};
\citealt{kir94}; \citealt{haf00}). However, the parameters of the bar,
e.g., its mass, size, and the motion of stars within it, still remain
poorly constrained.

\cite{sta97} used the Red Clump Giants (RCGs) to constrain the axial ratios
and orientation of the Galactic bar.  These stars are the equivalent of the
horizontal branch stars for a metal-rich population, i.e., relatively
low-mass core helium burning stars.  RCGs in the Galactic bulge occupy
a distinct region in the colour magnitude diagram (\citealt{sta00} and
references therein).  The intrinsic width of the luminosity distribution of
RCGs in the Galactic bulge is small, about 0.2 mag 
(\citealt{sta97}; \citealt{pac98}). Their observed peak and width of
the luminosity function are related to the distance and radial depth
of the bar.

Furthermore, \cite{mao02} suggested that the proper motion measurements
of RCGs in the Galactic center are useful in constraining the Galactic bar
parameters.  By considering a sub-sample of bright RCGs, one should be able
to isolate to a sufficient degree the stars that are on average closer to the
near side of the bar. Similarly, the stars in a faint sub-sample would be
more on the far side of the bar. If there is a tangential streaming motion of 100
km\,s$^{-1}$ in the bulge/bar, there should be a detectable difference
of 1.6 mas\,yr$^{-1}$ in the average proper motion between the bright and
faint RCG sub-samples. Measurements of this difference provide constraints
on the models of the Galactic bar.

To test the feasibility of the method, in this paper we analyze stellar
proper motions in one of the fields observed by the Optical Gravitational
Experiment II (OGLE-II; \citealt{uda00}).  In \S\,\ref{sec:data}
we describe the data.  We present the analysis method in
\S\,\ref{sec:analysis} and results in \S\,\ref{sec:results}.
Discussion and conclusion are given in \S\,\ref{sec:disc}.

\section{DATA}

\label{sec:data}

We use the data collected during the second phase of the
OGLE experiment, between 1997 and 2000. All observations were made
with the 1.3-m Warsaw telescope located at the Las Campanas Observatory,
Chile, which is operated by the Carnegie Institution of Washington.
The "first generation" camera has a SITe 2048 $\times$ 2048 pixel CCD detector
with pixel size of 24 $\mu$m resulting in 0.417 arcsec/pixel scale.
Images of the Galactic bulge were taken in drift-scan mode at "medium" readout
speed with the gain 7.1 $e^{-}$/ADU and readout noise of 6.3 $e^{-}$.
A single 2048 $\times$ 8192 pixel frame covers an area of 0.24 $\times$ 0.95
deg$^2$.  Saturation level is about 55,000 ADU.  Details of the
instrumentation setup can be seen in \cite{uda97}.

In this paper we use 266 $I$-band frames of the BUL\_SC1 field centered at
($\alpha, \delta)_{2000} =(18^h 02^m 32^s.5, -29^{\circ}57'41'')$.
The time baseline is almost 4 years.  There are gaps between the observing
seasons when the Galactic bulge cannot be observed from Las Campanas,
each about 3 months long.  The median seeing is $\sim 1.3''$.
We use the $VI$ photometric maps of standard OGLE template (\citealt{uda02}) 
as the astrometric and photometric references.

Only about 70\% of the area of the BUL\_SC1 field overlaps with the extinction
map made by \cite{sta96} and is used in the analysis. This ensures that we can
accurately deredden stellar magnitudes.

\section{Analysis}

\label{sec:analysis}

The standard OGLE template serves as the fixed astrometric 
reference in our analysis. In the case of BUL\_SC1 field, the frame adopted 
as the OGLE template was taken at JD = 2450561.715.  In order to treat properly
spatial PSF variations and frame distortions the field is divided
into 256 subframes before processing. Subframes are 512$\times$128 pixels
with 14 pixel margin on each side to smooth out transitions between
the local polynomial fits.  The shape of the subframe reflects stronger
y-axis (declination) gradients due to drift-scan mode of observation.
The actual proper motion analysis includes only 180 subframes,
that is 70\% of the area of the BUL\_SC1 field for which the accurate
interstellar extinction data is available from \cite{sta96}.

We compute the pixel positions of stars in each of the subframes 
using the DoPHOT
package (\citealt{sch93}).  At the start of the processing for each exposure,
the positions of stars in a single subframe are cross-referenced with those in
the template and the overall frame shift is obtained.  Using this crude shift
we can identify the same region of the sky (corresponding to a given subframe
of the template) throughout the entire sequence of frames.  For each of the
subframes, about $\sim 300$ brightest ($I<17$) stars categorized by DoPHOT as
isolated are used to derive the local transformation between pixel coordinate
systems of a given exposure and the template. The search radius in matching
the stars  between templete and other frames is 0.5 pixel.
 We use first order polynomial
to fit the transformation.  The resulting piece-wise transformation adequately
converts pixel positions to the reference frame of the template. Typical
residuals are at the level of $0.08$ pixels.

\begin{figure}
\begin{center}
\includegraphics[angle=-90,scale=0.37,keepaspectratio]{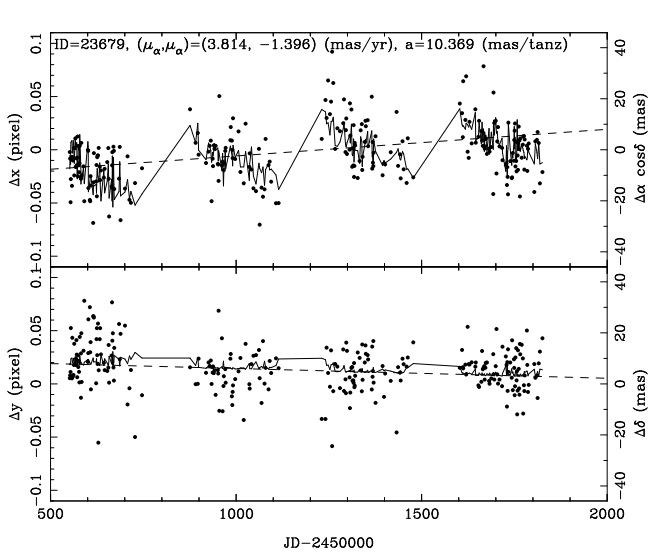}
\caption{Time variation of the position in $\alpha{\rm cos}\delta$ 
(upper) and $\delta$ (lower) of star ID=23679 (V-I=3.027).
Solid line indicates a model fit of the proper motion 
$(\mu_\alpha, \mu_\delta)=(3.8, -1.4)$ (mas\,yr$^{-1}$) with a 
differential refraction $a=10.4$ mas\,(tan$z$)$^{-1}$, and dashed 
line represents the same line without the term of differential 
refraction.
  \label{fig:shiftcurve}
  }
\end{center}
\end{figure}

An example of time dependence of the position for a star with the 
detectable proper
motion is shown in Fig. \ref{fig:shiftcurve}.  Also presented is the best fit
model of proper motion $(\mu_\alpha, \mu_\delta)$ with and without the
differential refraction.  The star's coordinates in the sky at the
time $t$ are given as follows:

\begin{equation}
  \label{eq:pm_model_a}
   \alpha = \alpha_0 + \mu_\alpha t + a\sin C\tan z, 
\end{equation}

\begin{equation}
  \label{eq:pm_model_d}
   \delta = \delta_0 + \mu_\delta t + a\cos C\tan z, 
\end{equation}
where $a$, $z$ and $C$ are the differential refraction coefficient,
the zenith angle and the angle made by the Zenith, the star and the South
Pole; $\alpha_0$ and $\delta_0$ are constants. The parameter $a$
is a function of the apparent star colour.  Here we neglect the parallax
effect due to the Earth motion because we are interested in stars
at the distance of the Galactic center.

We computed $\alpha_0$, $\delta_0$, $\mu_\alpha$, $\mu_\delta$ and $a$
for all 46,961 stars used to transform coordinate systems (approximately
the number of used subframes times the typical number of stars per subframe,
180 $\times$ 300).  In cases when the star is measured in the overlap region
of more than one subframe, the data set with the largest number of points is
selected.  Stars with the number of data points fewer than 20 are rejected.

A sample of fitted proper motions is listed in Table \ref{tbl:list}.  The
complete list of all 46,961 stars is available in electric format via
anonymous ftp from the server {\it astro.princeton.edu}, directory {\it
/sumi/propermotion/bul\_sc1.pm.gz}.  The list contains star ID, number 
of data points, measured $\mu_\alpha$ and $\mu_\delta$ with their errors, $a$, 
standard deviation (Sdev) of data points in the fitting, equatorial 
coordinates, apparent $I$-band magnitude and $V-I$ colour, and extinction 
corrected magnitude $I_0$ and colour $(V-I)_0$ estimated using the 
extinction map of \cite{sta96}. ID, coordinates, $I$ and $V-I$ for each 
object are identical with those in \cite{uda02}.

\begin{table*}
 \centering
 \caption{Sample of fitted parameters. 
    \label{tbl:list}}
    \begin{tabular}{lcrcrcrrcccccc}\\
    ID &N& $\mu_\alpha \cos\delta$ &$\sigma_{\mu_{\alpha\cos\delta}}$& $\mu_\delta$ & 
     $\sigma_{\mu_\delta}$ & $a$  & Sdev & $\alpha_{2000}$ & $\delta_{2000}$ 
        & $I$ & $V-I$ & $I_0$ & $(V-I)_0$\\
     && \multicolumn{4}{c}{(mas\,yr$^{-1}$)} & \multicolumn{2}{c}{(mas)} &
      (hour)&(deg.)&  \multicolumn{4}{c}{(mag)} \\
  \hline
 11933&263&-3.86&0.49& 5.68&0.49&  0.70& 9.45&18.0355861&-30.3390194&12.617&1.878&11.638&1.241\\
 11934&230&-3.18&0.52& 4.45&0.52&-10.40& 9.37&18.0359096&-30.3321479&12.254&0.663&11.195&-0.022\\
 11935&263& 1.07&0.58& 1.95&0.58&  3.01&11.34&18.0348306&-30.3262084&12.747&2.257&11.667&1.550\\
 11936&262&-4.46&0.46&-0.13&0.45& 34.05& 8.80&18.0358950&-30.3248649&12.868&4.805&11.790&4.092\\
 11937&243&-1.24&0.51& 1.12&0.51& 31.46& 9.47&18.0346639&-30.3174773&12.503&5.206&11.400&4.494\\
 11938&246& 1.19&0.51& 3.11&0.51& 20.69& 9.53&18.0349745&-30.3175164&12.492&4.406&11.390&3.690\\
 11939&198&-4.30&0.71& 0.76&0.70& 14.21&11.82&18.0351497&-30.3174968&12.247&3.679&11.145&2.961\\
 11943&261& 1.84&0.54& 3.73&0.54&-11.53&10.50&18.0349116&-30.3700535&13.220&0.483&12.241&-0.154\\
 11945&262&-1.43&0.45&-3.92&0.45&  3.06& 8.79&18.0345371&-30.3657689&13.225&2.597&12.276&1.979\\
 11948&259&-1.10&0.41& 0.02&0.41& 26.13& 7.83&18.0345576&-30.3619234&13.072&4.608&12.129&3.990\\
 11950&266& 0.28&0.53& 2.26&0.53&  4.35&10.33&18.0367544&-30.3590652&13.490&2.135&12.500&1.509\\
 11951&265&-4.08&0.53& 1.23&0.53& 10.62&10.30&18.0365689&-30.3574343&13.416&3.648&12.395&2.999\\
 11952&248& 1.59&0.54&-6.59&0.54&  6.22&10.10&18.0342311&-30.3572907&13.157&3.256&12.203&2.626\\
 11953&244&-8.37&1.42&-2.66&1.41& -3.02&26.65&18.0364769&-30.3556208&13.294&0.671&12.243&0.003\\
 11955&265&-0.56&0.37& 0.06&0.37&  4.83& 7.25&18.0354601&-30.3520741&13.700&2.446&12.708&1.811\\
\end{tabular}
\end{table*}

\begin{figure}
\begin{center}
\includegraphics[angle=0,scale=0.45,keepaspectratio]{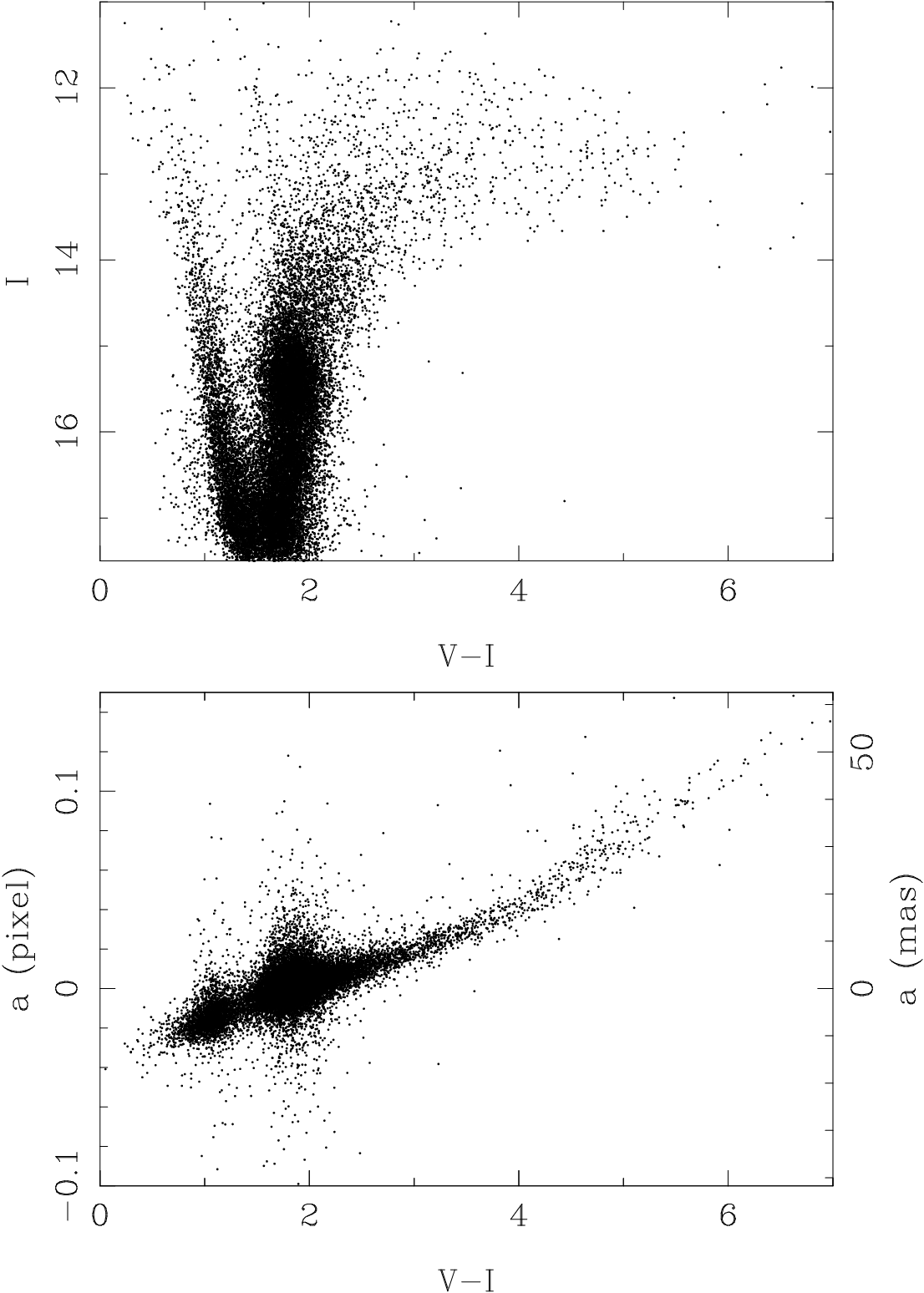}
\caption{Upper: Colour magnitude diagram of the half of the analysed stars.
Lower: Correlation between the differential refraction
coefficient $a$ and the apparent colour (uncorrected for extinction)
for stars with $I<16$.
  \label{fig:difref}
  }
\end{center}
\end{figure}

\begin{figure}
\begin{center}
\includegraphics[angle=-90,scale=0.35,keepaspectratio]{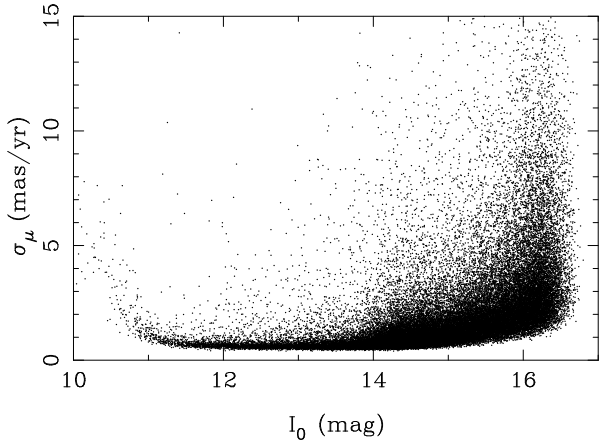}
\caption{Uncertainty in $\mu$ as a function of the extinction corrected I-band
magnitude $I_0$.
  \label{fig:sigma}
  }
\end{center}
\end{figure}

In Fig. \ref{fig:difref}, we present the $I$ and $V-I$ Colour Magnitude
Diagram (CMD) of stars used in this analysis. We also show the correlation
between the differential refraction coefficient $a$ and the apparent
$V-I$ colour (uncorrected for extinction) for stars with $I<16$. 
In Fig. \ref{fig:sigma}, we plot the uncertainty in $\mu$, $\sigma_\mu$, as a function
of extinction corrected $I$-band magnitude $I_0$, and we list the mean $<\sigma_\mu>$
in Table \ref{tbl:sigma}.

\begin{table}
 \centering
 \caption{Mean of uncertainty in $\mu$, $<\sigma_\mu>$ (mas\,yr$^{-1}$),
    as a function of $I_0$ (mag). $\sigma_\mu$ is averaged over $I_0\pm0.5$.
    \label{tbl:sigma}}
    \begin{tabular}{cccccccc}\\
$I_0$          & 10   & 11  & 12  & 13  & 14  & 15 & 16 \\
$<\sigma_\mu>$ & 4.84 & 1.40& 0.83& 0.88& 1.18& 1.80&3.44\\
\end{tabular}
\end{table}

\begin{figure}
\begin{center}
\includegraphics[angle=-90,scale=0.35,keepaspectratio]{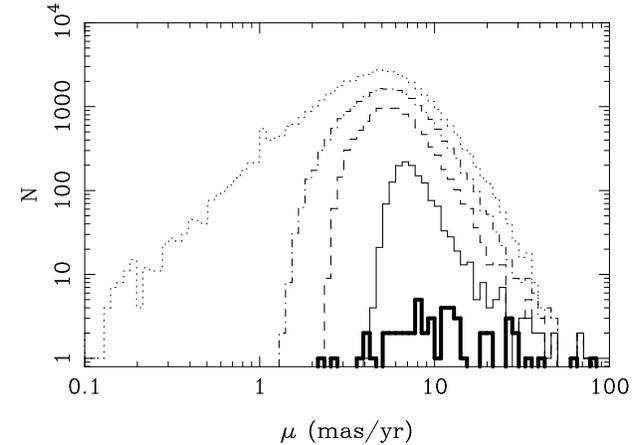}
\caption{Histogram of $\mu_{\rm EW}$ (thick solid line), $\mu$ of all 46,961 sample
(dotted line), that detected with confidence better than $3\sigma$ (dot-dashed line),
better than $5\sigma$ (dashed line) and better than $10\sigma$ (thin solid line).
  \label{fig:hpmhist}
}
\end{center}
\end{figure}

To check the measurements we cross-identified our stars with the list of high
proper motion objects detected photometrically by \citet{eye01}.  Out of 74
stars in \citet{eye01}, 53 are in the region used in the present analysis,
and 52 were recovered.  We plot the proper motions of those 52 stars as
measured in both \citet{eye01} ($\mu_{\rm EW}$) and this work ($\mu$).
The very good correlation gives us certain confidence in our measurements. 
Fig. \ref{fig:hpmhist} shows the histogram of $\mu_{\rm EW}$ (thick solid line), 
$\mu$ of all 46,961 sample (dotted line), the sample detected with the
confidence better than $3\sigma$ (dot-dashed line), better than $5\sigma$
(dashed line) and better than $10\sigma$ (thin solid line).
We can see that our method is more effective in detecting proper motions 
than \citet{eye01}.

\begin{figure}
\begin{center}
\includegraphics[angle=-90,scale=0.35,keepaspectratio]{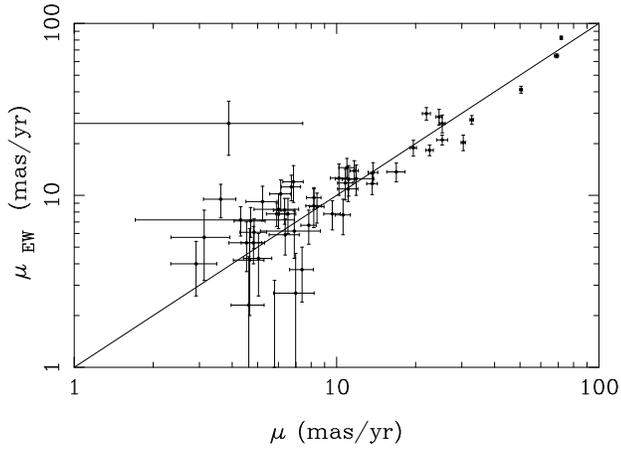}
\caption{Comparison of proper motions for 52 stars measured in both 
  \citet{eye01} ($\mu_{\rm EW}$) and this analysis ($\mu$).
  \label{fig:crossref}
}
\end{center}
\end{figure}

\section{Results}
\label{sec:results}

\begin{figure}
\begin{center}
\includegraphics[angle=0,scale=0.45,keepaspectratio]{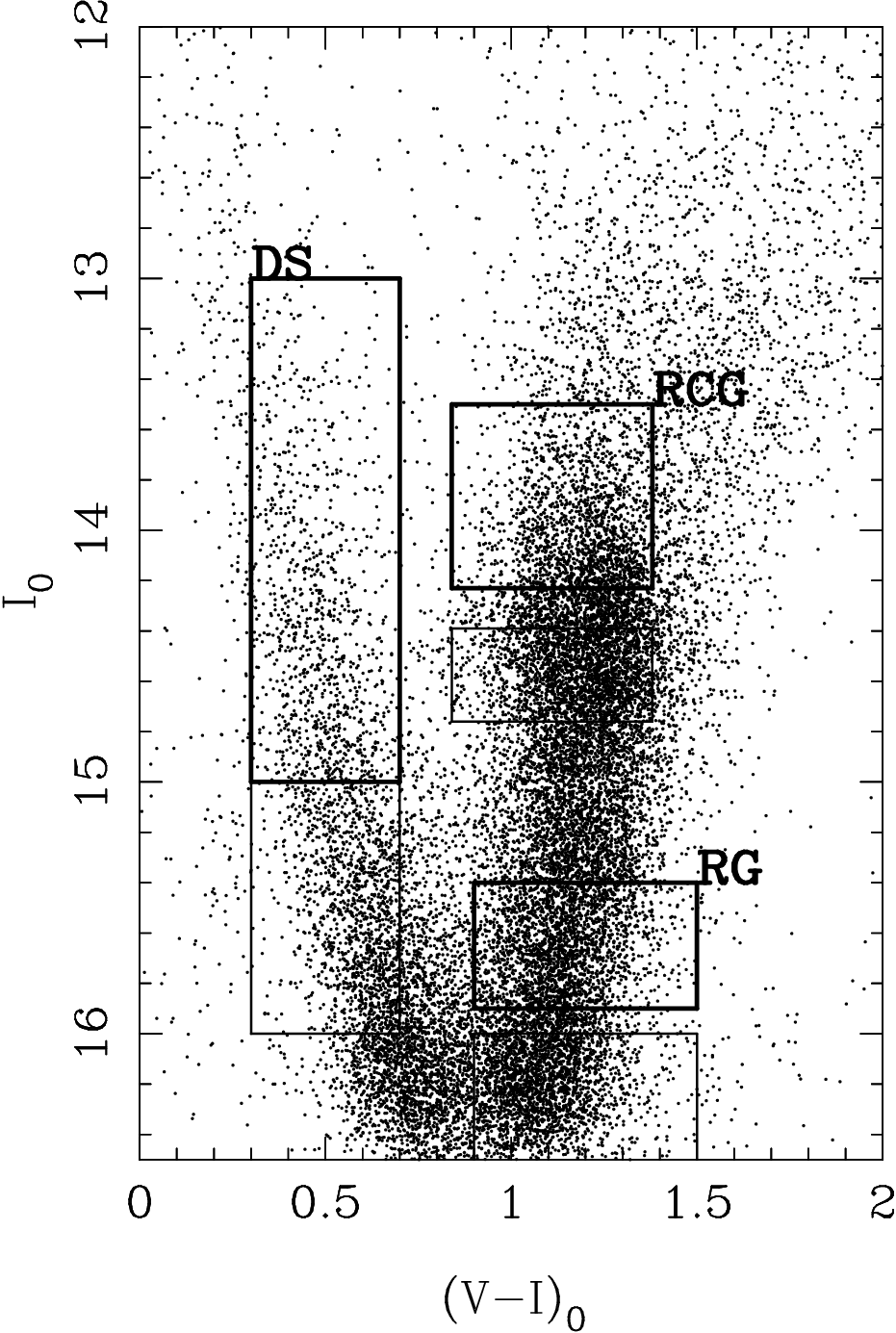}
\caption{Extinction corrected Colour Magnitude diagram of the half of stars
used in this analysis. Three populations RCGs, Disc Stars (DSs) 
and Red Giants (RGs) are enclosed by upper-right, upper-left and 
lower-right boxes. Thick and thin boxes represent bright and faint 
samples, respectively.
  \label{fig:CMD}
  }
\end{center}
\end{figure}

For the purpose of proper motion analysis we identify three populations
of stars: RCGs, Disc Stars (DSs) and Red Giants (RGs).  Those occupy
respectively the upper-right, upper-left and lower-right regions in
Fig. \ref{fig:CMD}, the extinction corrected CMD.
Within each population we distinguish a bright and a faint sub-sample
indicated by thick and thin boxes in the same figure.  Those regions
of the CMD are defined by the following formula:

\begin{equation}
  \label{eq:RC_Ib}
    13.5  <I_0< 14.23 \quad ({\rm bright\, RCGs}),
\end{equation}

\begin{equation}
  \label{eq:RC_If}
 14.39   <I_0< 14.76  \quad ({\rm faint\, RCGs}),
\end{equation}

\begin{equation}
  \label{eq:RC_VI}
   0.84   <(V-I)_0< 1.38 \quad ({\rm both\, RCGs}),
\end{equation}
\begin{equation}
  \label{eq:DS_Ib}
    13.0  <I_0< 15.0 \quad ({\rm bright\, DSs}),
\end{equation}

\begin{equation}
  \label{eq:DS_If}
 15.0  <I_0< 16.0 \quad ({\rm faint\, DSs}),
\end{equation}

\begin{equation}
  \label{eq:DS_VI}
   0.3   <(V-I)_0< 0.7 \quad ({\rm both\, DSs}),
\end{equation}
\begin{equation}
  \label{eq:RG_Ib}
    15.4  <I_0< 15.9 \quad ({\rm bright\, RGs}),
\end{equation}

\begin{equation}
  \label{eq:RG_If}
 16.0   <I_0< 16.5 \quad ({\rm faint\, RGs}),
\end{equation}

\begin{equation}
  \label{eq:RG_VI}
   0.9   <(V-I)_0< 1.5 \quad ({\rm both\, RGs}).
\end{equation}

The CMD region for the RCGs was found to maximize the significance 
of the proper motion difference between bright and faint samples. 

\begin{figure}
\begin{center}
\includegraphics[angle=-90,scale=0.31,keepaspectratio]{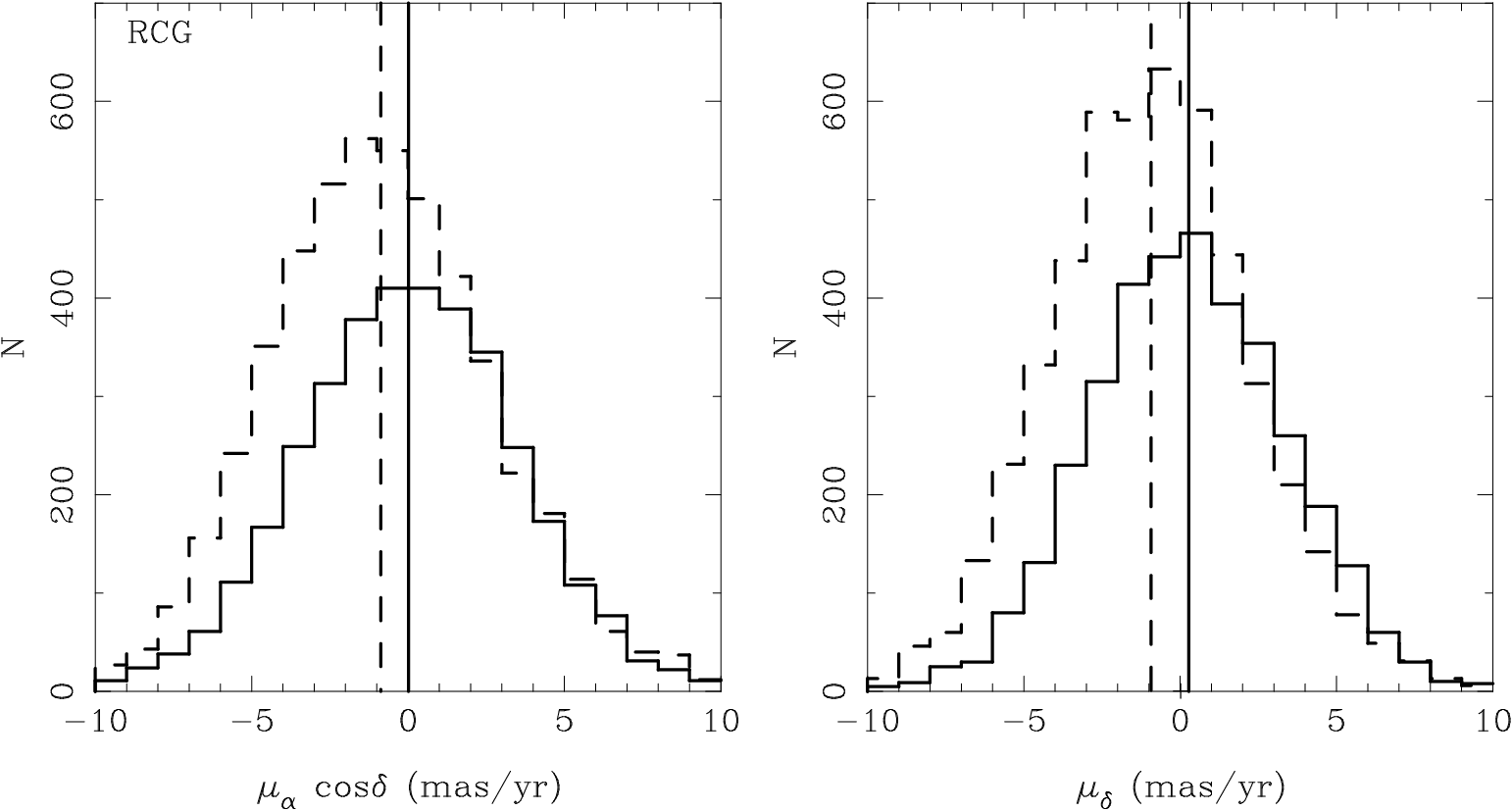}
\includegraphics[angle=-90,scale=0.31,keepaspectratio]{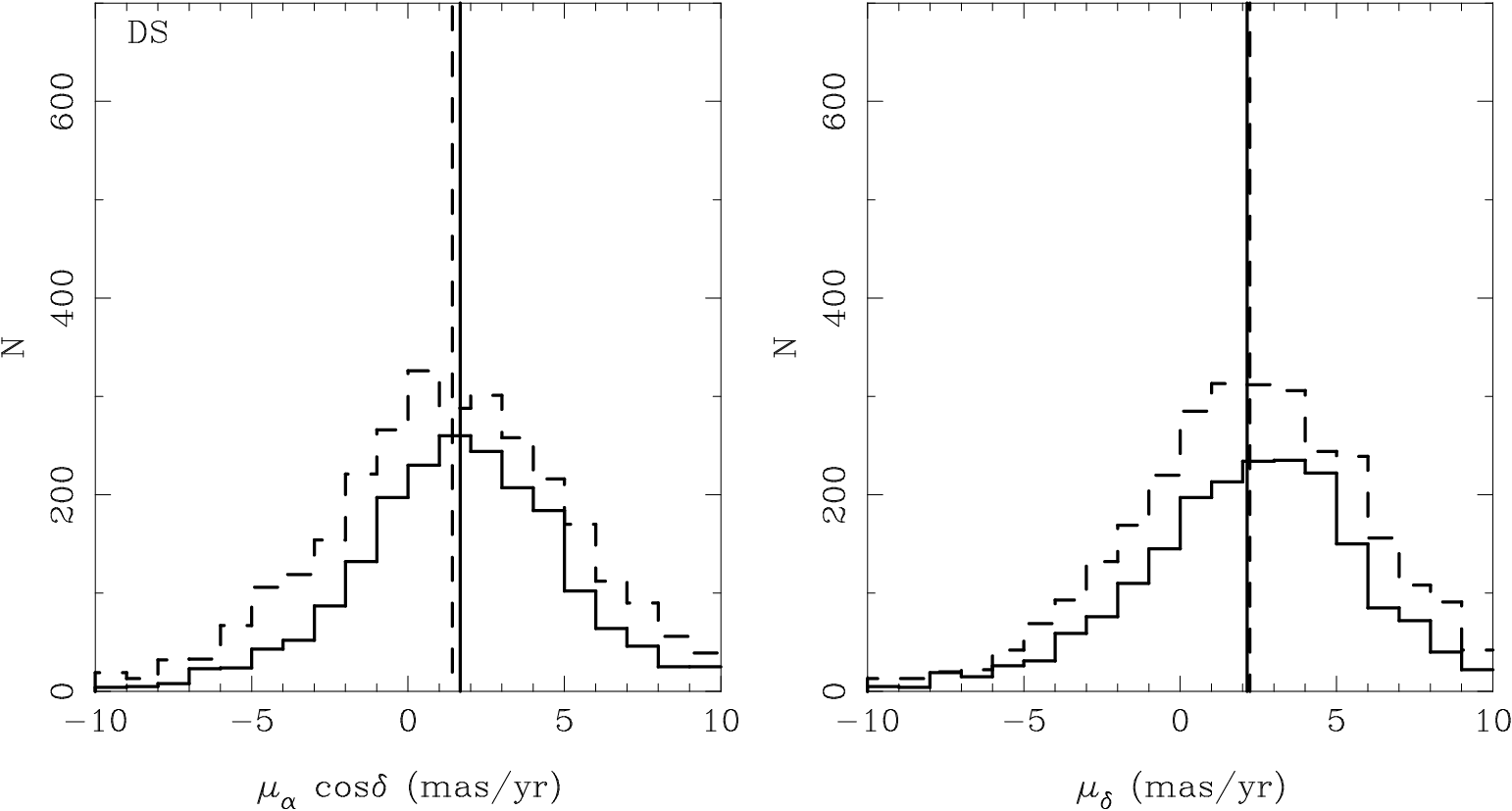}
\includegraphics[angle=-90,scale=0.31,keepaspectratio]{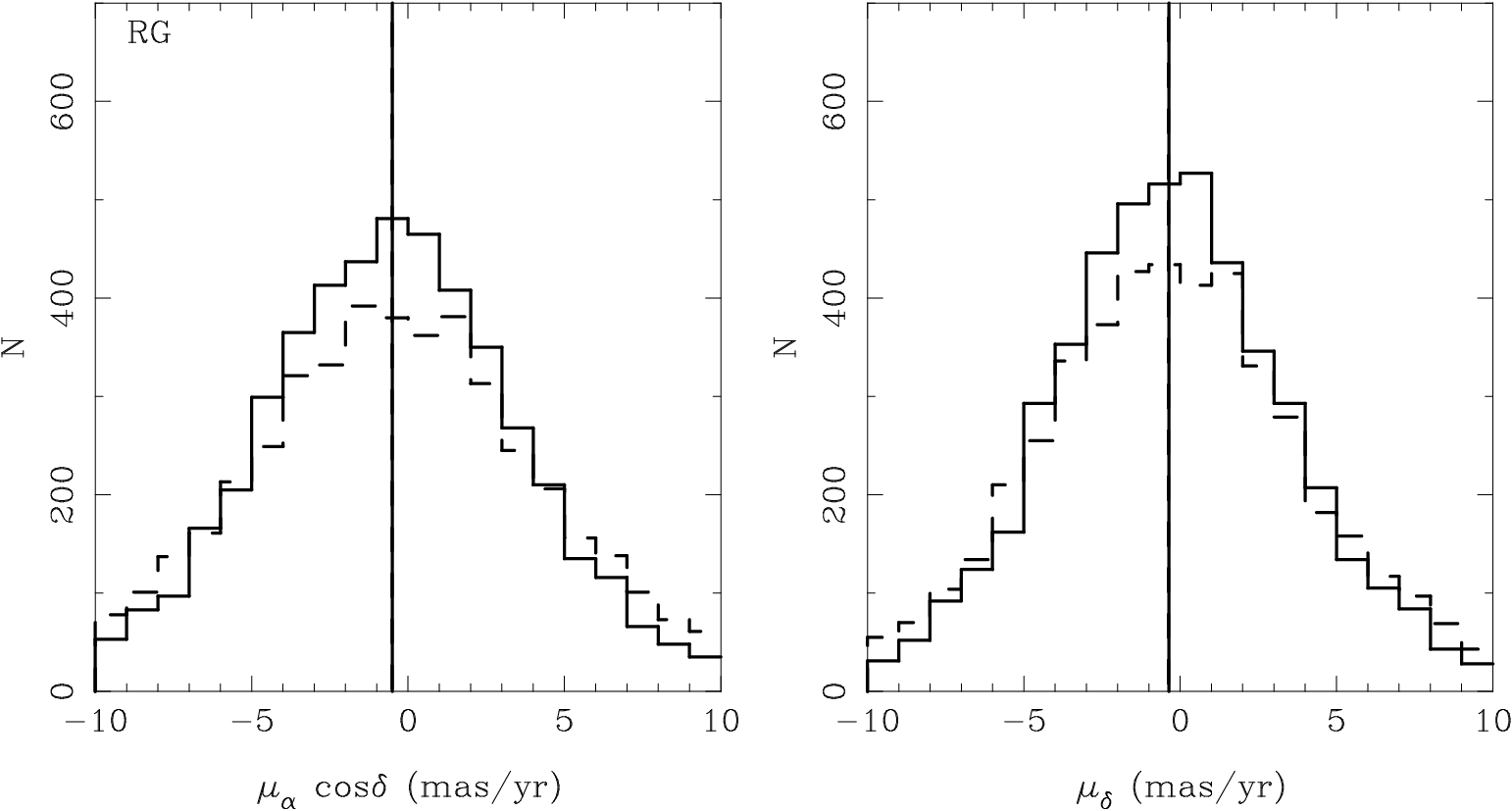}
\caption{Histogram of the distribution of $\mu_\alpha$ (left)
and $\mu_\delta$ (right) for bright (thick line) and faint (dashed line) 
samples for populations RCGs, DSs and RGs. 
  \label{fig:hist_pm}
  }
\end{center}
\end{figure}

\begin{figure}
\begin{center}
\includegraphics[angle=-90,scale=0.42,keepaspectratio]{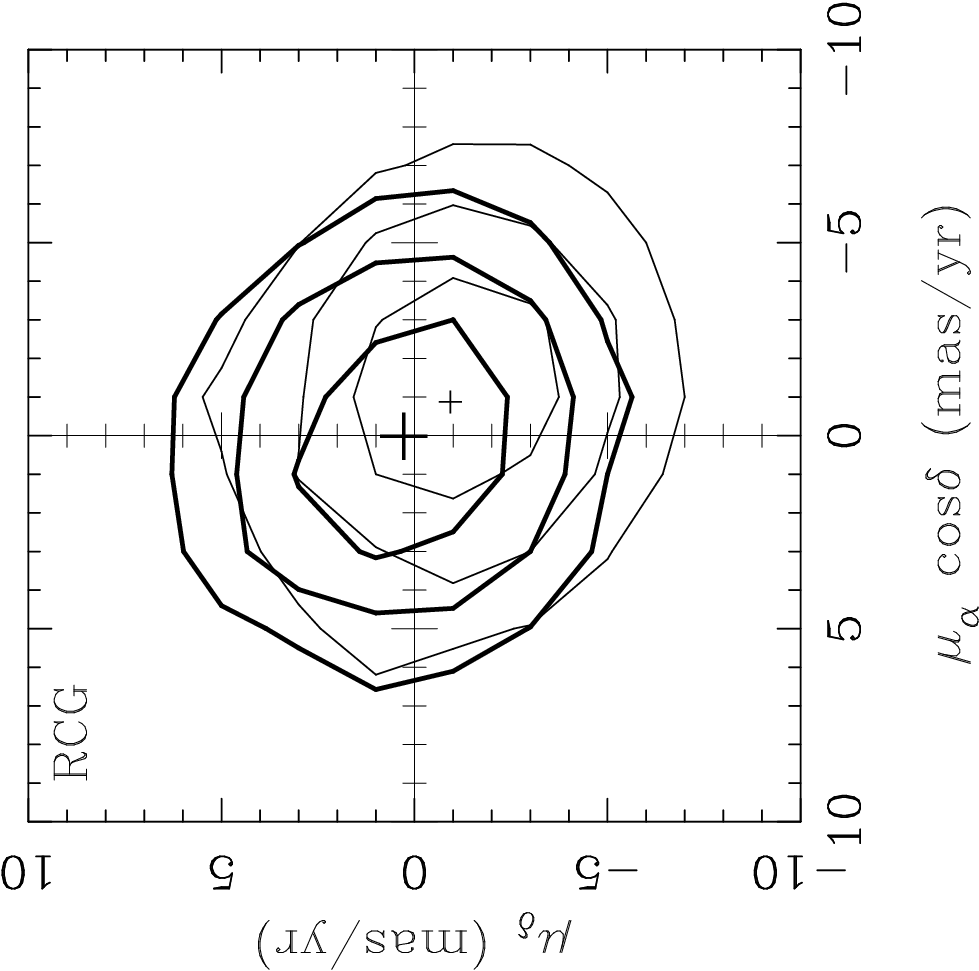}
\includegraphics[angle=-90,scale=0.42,keepaspectratio]{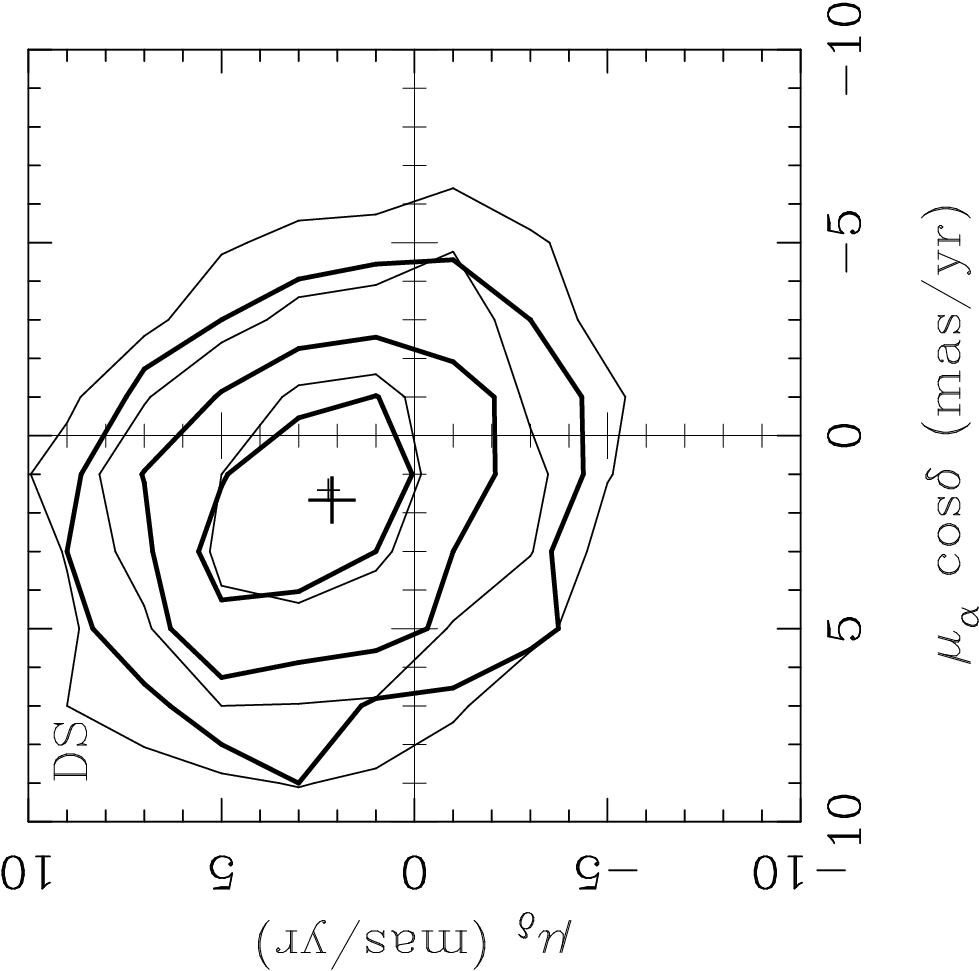}
\includegraphics[angle=-90,scale=0.42,keepaspectratio]{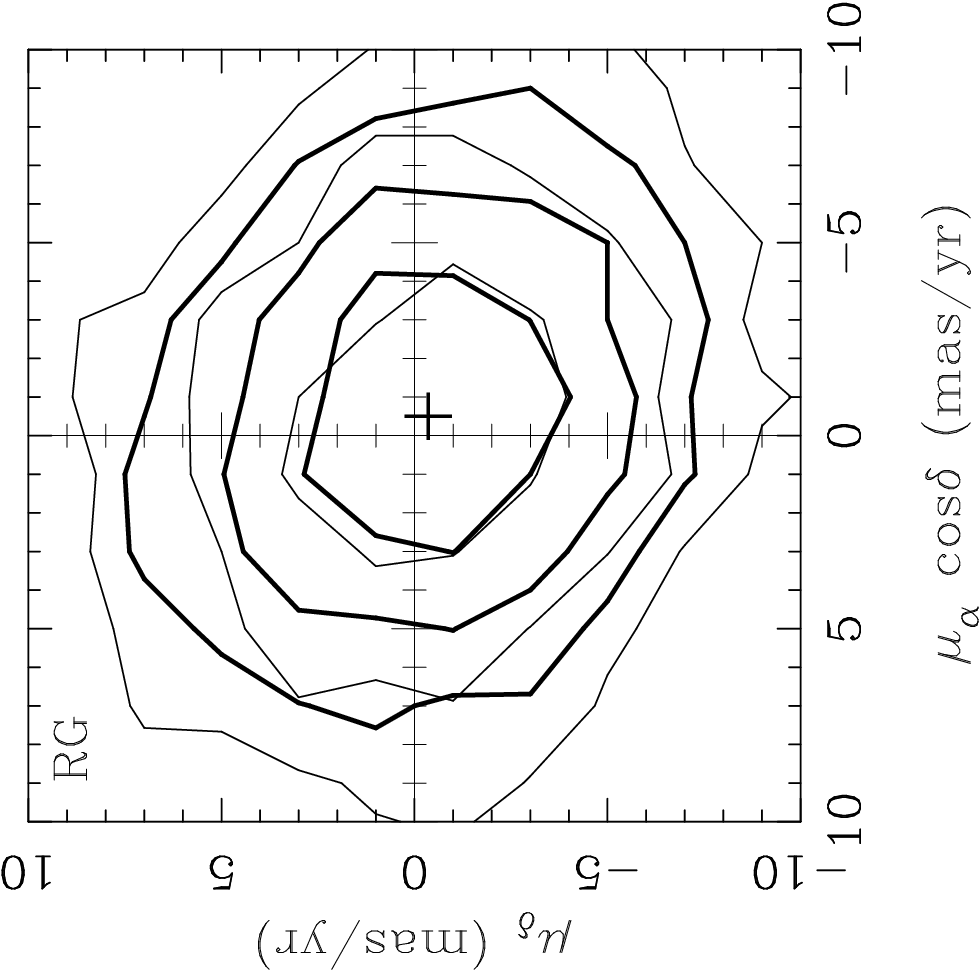}
\caption{Contour map and mean of the distribution of $\mu_\alpha$ and 
$\mu_\delta$ for bright (thick line and cross) and faint (thin line and 
cross) samples within three stellar populations defined in the analysis:
RCGs (Top), DSs (Middle) and RGs (Bottom). Contours enclose 30, 60 and 80\% of stars.
Top-left corner corresponds to positive $l$ and bottom-right corner to negative 
$l$ direction.
  \label{fig:contour_pm}
  }
\end{center}
\end{figure}

Figs. \ref{fig:hist_pm} and \ref{fig:contour_pm} summarize the main results.
In Fig. \ref{fig:hist_pm} we show a histogram of $\mu_\alpha$ (left panel)
and $\mu_\delta$ (right panel) for bright (thick line) and faint (dashed line)
samples of all three stellar populations.  Fig. \ref{fig:contour_pm} shows the contour
map and the mean of the distribution in $\mu_\alpha$ and $\mu_\delta$ 
for bright (thick line and cross) and faint (thin line and cross) samples 
of RCGs, DSs and RGs. In this figure, the top-left corner is toward positive $l$
and bottom-right corner is toward negative $l$ direction.  We also show the
mean value of the relative proper motion $<\mu>$ in equatorial and Galactic
coordinates as well as the corresponding Root Mean Square (RMS)
in equatorial coordinates for each of the samples in Table \ref{tbl:mean_mu}.
We assumed that the error in $\mu$ of each star is the RMS of the related distribution. 
This RMS is the combination of the intrinsic scatter in $\mu$ and the
astrometric uncertainty and therefore it provides a sensible upper limit for the
error in $<\mu>$. In measuring the mean value we rejected high proper 
motion objects with $|\mu| >20$ mas\,yr$^{-1}$.
Note that the mean values $<\mu>$ of all sample are not exactly zero, 
as these are not identical to all stars used to align the
frames. Stars with fewer than 20 data points and detected multiple times in
the overlapping subframes were rejected. The differences are insignificant
and we can safely ignore them.

\begin{figure}
\begin{center}
\includegraphics[angle=-90,scale=0.32,keepaspectratio]{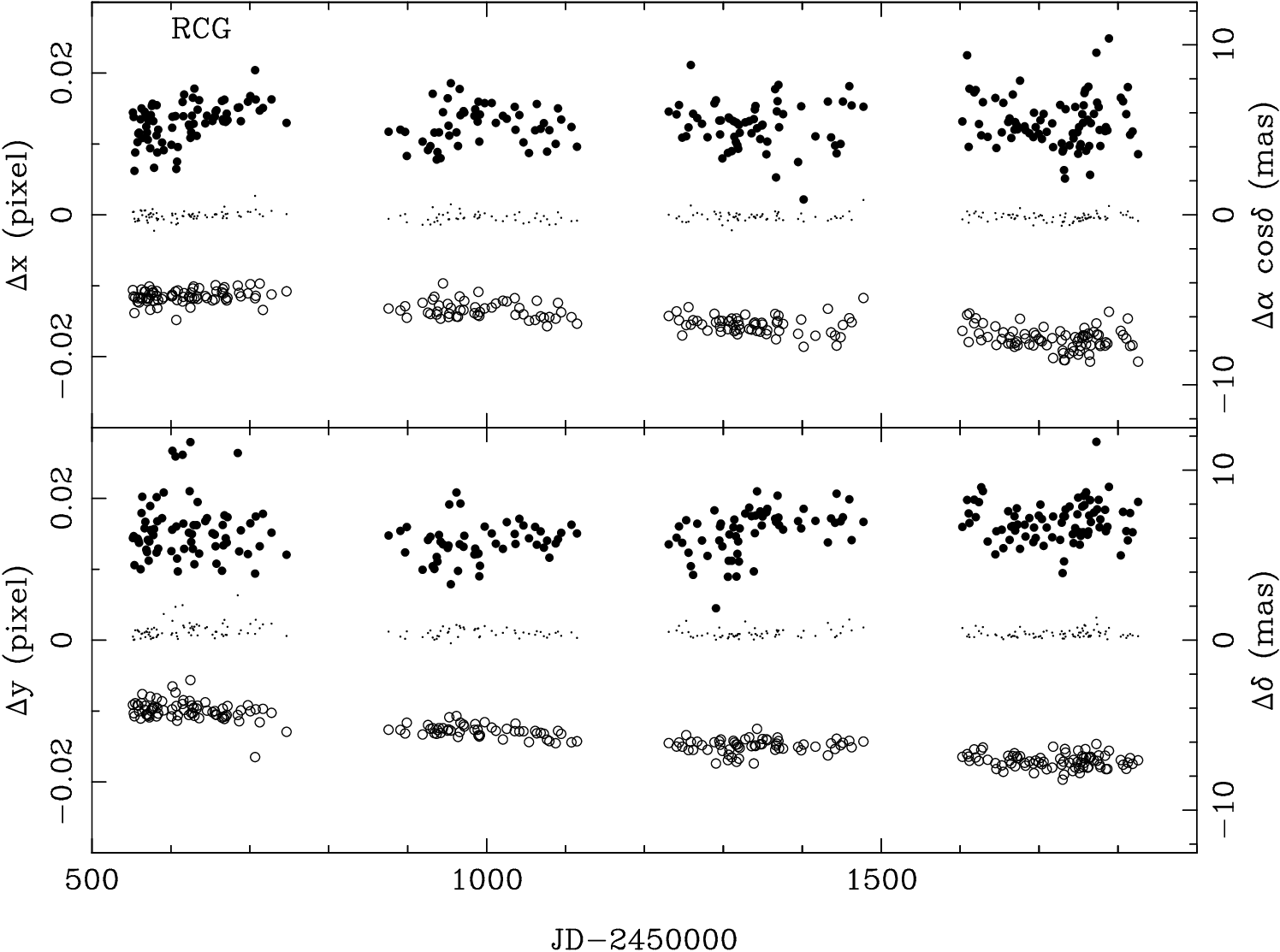}
\includegraphics[angle=-90,scale=0.32,keepaspectratio]{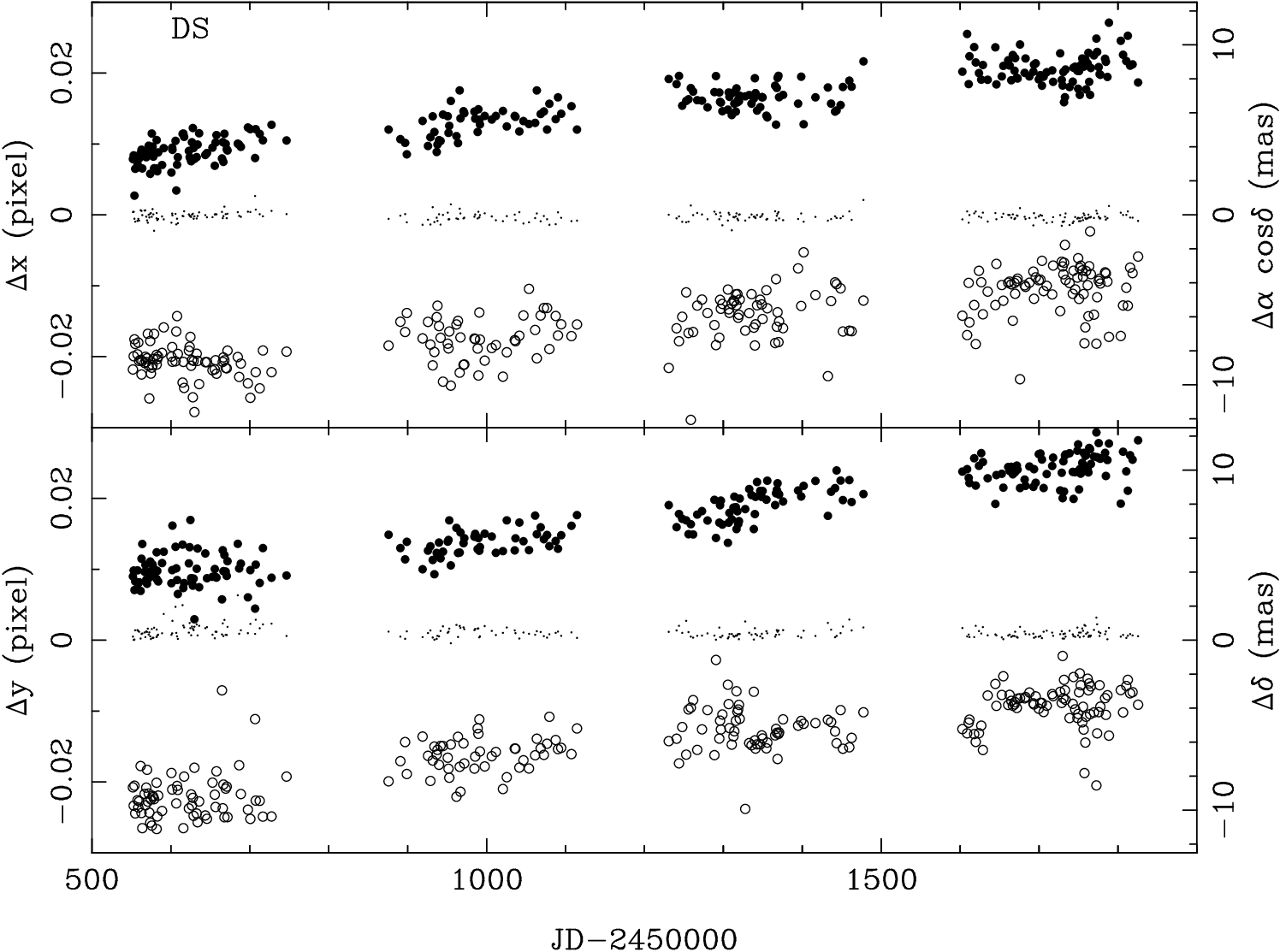}
\includegraphics[angle=-90,scale=0.32,keepaspectratio]{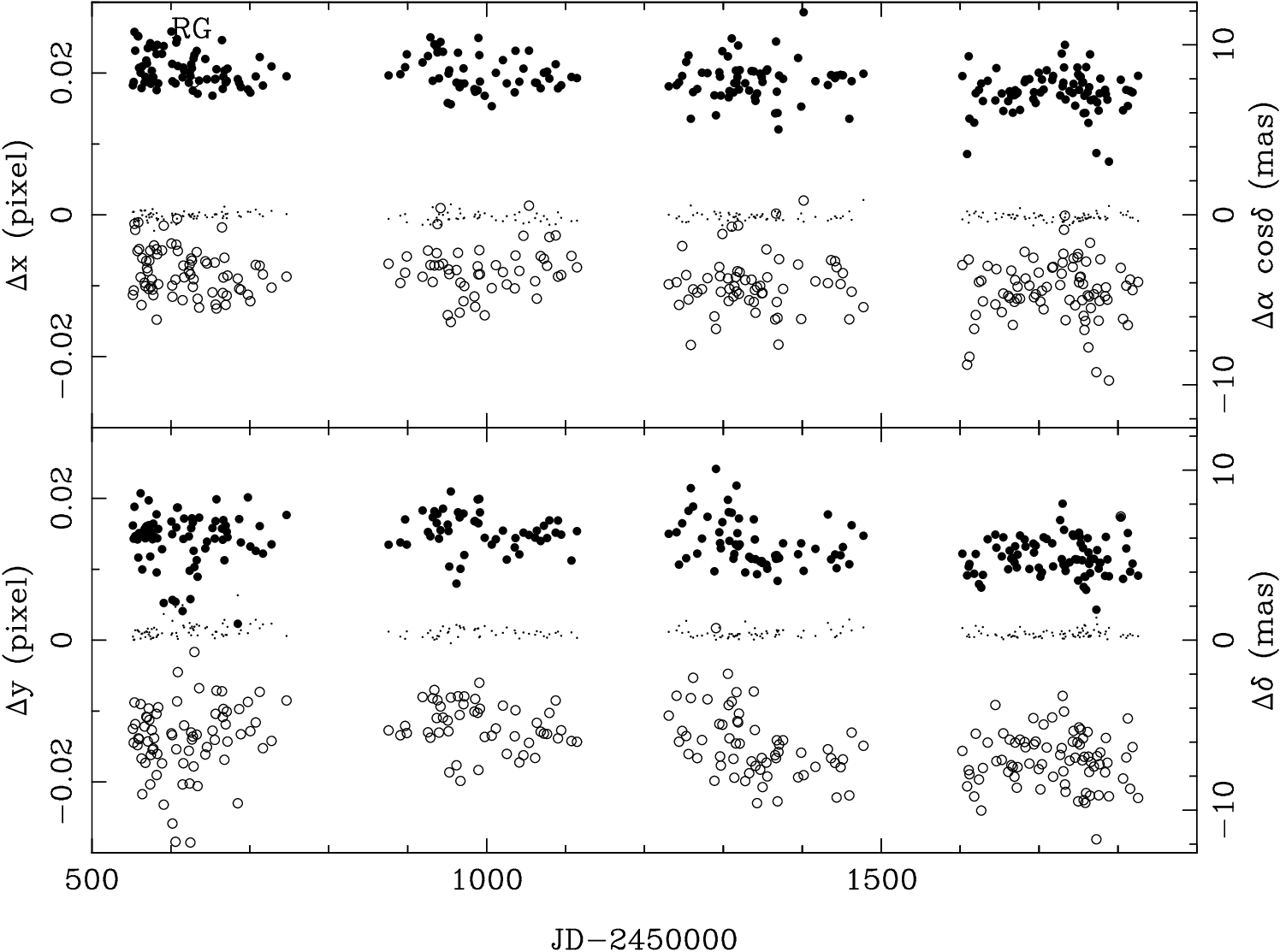}
\caption{The mean positional shift as a function of JD for RCGs (Top), 
DSs (Middle) and RGs (Bottom).  The data points for each of the samples are
shifted vertically for clarity and shown with different symbols:
all (dot), bright (filled circle) and faint
(open circle).
  \label{fig:shift_time}
  }\end{center}
\end{figure}

Finally, we also show the mean positional shift of stars in each
population as a function of time (JD) in Fig. \ref{fig:shift_time}.
The plots of each samples, i.e., all (dot), bright (filled circle)
and faint (open circle), are shifted vertically for clarity,
because only the slope is important.

In Figs. \ref{fig:hist_pm}, \ref{fig:contour_pm} and Table \ref{tbl:mean_mu},
differences in $<\mu>$ between bright and faint samples of RCGs are clearly
seen. The main component is along $<\mu_l>$ and reaches $\Delta <\mu> \sim 1.5$
mas\,yr$^{-1}$. The accuracy in $\Delta <\mu>$ is about $\sigma \sim 0.11$
mas\,yr$^{-1}$, therefore the significance of the difference is $\sim 14 \sigma$.
In short, stars at the far side are moving toward negative 
Galactic latitude $l$ relative to the stars at the near side.  
The magnitude and orientation of the proper motion of one group relative 
to the other can be explained by  the rotation of stars in the bar in the 
same direction as the Solar motion around the Galactic center. The result is 
consistent with $\Delta <\mu> \sim 1.6$
mas\,yr$^{-1}$ estimated by \cite{mao02} using the \cite{kir94}'s
model with the tangential streaming motion of $v_b \sim 100$ km\,s$^{-1}$.
Note that in this analysis the measured $<\mu>$ is not absolute, but 
relative. Although the astrometric reference frame seems to be close to the one
based on stars at the near side of the bar, in principle it is not associated
with any pneumatically defined stellar population. On the other hand,
the difference $\Delta <\mu>$ between the two populations of RCGs is properly
defined and reliable.


\begin{table*}
 \centering
 \caption{Mean values, errors and RMS of $\mu$ (mas\,yr$^{-1}$) for star
 samples discussed in the analysis.
    \label{tbl:mean_mu}}
    \begin{tabular}{lrcccccccccc}\\
    Population &N& $<\mu_\alpha cos\delta>$ &RMS& $<\mu_\delta>$ & RMS 
                 & $<\mu_l >$ & $<\mu_b>$\\
  \hline
all       &46961&-0.046$\pm$0.022&4.69&0.091$\pm$0.020&4.39&0.054$\pm$0.021&0.086$\pm$0.021\\
bright RCGs&3610& 0.012$\pm$0.060&3.60&0.279$\pm$0.055&3.29&0.245$\pm$0.056&0.135$\pm$0.059\\
faint  RCGs&4979&-0.871$\pm$0.053&3.76&-0.930$\pm$0.048&3.37&-1.247$\pm$0.049&0.259$\pm$0.052\\
bright DSs& 2004&1.669$\pm$0.082&3.67& 2.140$\pm$0.084&3.78& 2.697$\pm$0.084& -0.310$\pm$0.083\\
faint  DSs& 3011&1.413$\pm$0.081&4.47& 2.229$\pm$0.078&4.26& 2.639$\pm$0.079& -0.045$\pm$0.080\\
bright RGs& 4911&-0.501$\pm$0.067&4.68& -0.356$\pm$0.060&4.24& -0.565$\pm$0.062&0.242$\pm$0.065\\
faint  RGs& 4770&-0.516$\pm$0.080&5.55& -0.378$\pm$0.072&4.97& -0.592$\pm$0.074&0.244$\pm$0.078\\

\end{tabular}
\end{table*}

Both RG samples have same proper motion and values of $<\mu_l>$ 
are about a mid between the those of bright and faint RCGs. This can be
explained by that RGs are on average at the center of the bar.

The disc stars seem to be moving toward positive $l$ with $<\mu_l>
\sim 3.3$ mas\,yr$^{-1}$ with respect to RGs, i.e., the Galactic center. 
This is roughly consistent with $<\mu_l> \sim 3.2$ mas\,yr$^{-1}$ estimated 
assuming that disc stars are on average on the $\sim 1$ kpc 
from the Sun and the flat rotation curve of 
$v_r \sim 220$ km\,s$^{-1}$. Here we also assumed that the Sun has 
an additional velocity of 19 km\,s$^{-1}$ toward ($l,b)=(53^\circ, 25^\circ$)
relative to the Local Standard of Rest (RSL) (\citealt{bin87}), i.e., 
$(v_{\odot l},v_{\odot b})$= (12 km\,s$^{-1}$, 7 km\,s$^{-1}$).
As for the disc stars, their distance may be estimated from their
apparent brightness.  They are typically at the main sequence
turn-off point for an old population, so very crudely they have 
absolute magnitude $M_{I,0} \approx 4$ or so.  The bright DSs are
at about $I_0$ = 14 mag, the faint are at about $I_0$ = 15.5 mag.
So, their distance moduli are approximately 10 mag and 11.5 mag,
which corresponds to the distance from us: 1 kpc and 2 kpc,
respectively. For the faint DSs, slightly larger proper motion
with respect to RGs is expected than the measured one.
This difference might be due to the difference in rotation curve along
the line of sight. 
The values $<\mu_b> \sim -0.3$ (bright DSs) or $-0.05$ (faint DSs) 
might be explained by the solar motion $(v_{\odot l},v_{\odot b})$ 
relative to LSR.  A detailed analysis of these results is beyond
the scope of this pilot study.

\section{Discussion and Conclusion}

\label{sec:disc}

We have measured the proper motion for 46,961 stars in the OGLE-II
BUL\_SC1 field covering Baade's window. We dramatically increased
the number of objects with large proper motions in this field compared
to \cite{eye01} who detected 53 high proper motion objects in the same region 
of the Galactic bulge.  We have estimated the difference in the proper 
motion between the bright (near-side) and the faint (far-side) samples 
of RCGs. We found that the far-side RCGs have a proper motion of 
$\Delta <\mu> \sim 1.5 \pm 0.011$ mas\,yr$^{-1}$ toward the negative 
$l$ relative to the near-side RCGs.  The results fit the picture with 
stars in the bar rotating around the Galactic center in the same 
direction as the Sun.  The value $\Delta <\mu> \sim$ 1.5 mas\,yr$^{-1}$ is 
consistent with 1.6 mas\,yr$^{-1}$ estimated by \cite{mao02} who
assumed a streaming motion of the bar at $v_b \sim 100$ $km\,s^{-1}$.
The presented method used with the OGLE-II data is sensitive to 
the relative streaming motion of stars in the Galactic bar
down to about $\Delta <\mu> \sim 0.1$ mas\,yr$^{-1}$.

As  discussed  in  the  previous  section,  the measured proper motions
of DS and RG samples seem consistent with the basic understanding of
stellar motions in the Galaxy, and the Solar motion relative to LSR in
particular. This consistency supports the reliability of our analysis.

The question of possible contamination by stars in the Galactic disc
bears some discussion as we select the samples of RCGs using
the CMD. Only nearby disc main-sequence stars, or evolved disc stars
are expected to fall in the same CMD region as our RCG samples.
The effect can be considered negligible because these stars are not
very numerous (\citealt{sta94}).

One should keep in mind that all measurements of $<\mu>$ presented here
are not absolute, but relative to the astrometric reference frame
which is not well known. This problem can be solved 
by using background quasars that would be detected in the near future
using the OGLE-II variability catalog (\citealt{woz02,eye02}).
The primary goal of this paper is to demonstrate that proper motions
can be measured with very high precision using the OGLE-II data,
sufficiently high to clearly detect the presence of a strong streaming motion
(rotation) of stars in the Galactic bar.  While the reference frame established by
the large number of stars is not well defined with respect to the inertial
frame of reference, the relative motions of groups of stars are well
determined.  A detailed analysis of these results is beyond the scope
of the present study. We intend to expand our work to all 49 OGLE-II
Galactic bulge fields covering a large range of $l$ and $b$ around the bar.
A thorough analysis of all available data is underway and will be published
elsewhere.

\section*{Acknowledgments}

We are grateful to B. Paczy\'{n}ski for helpful comments and
discussions. 
We are grateful to the OGLE team for providing us with all CCD images on
which this paper is based.
T.S. and L.E. acknowledge the financial support from the
Nishina Memorial Foundation and from the Swiss National Science
Foundation respectively. 
This work was partly supported with the following grants to B. Paczy\'nski:
NSF grants AST-9820314 and AST-0204908, and NASA grants NAG5-12212,
and grant HST-AR-09518.01A provided by NASA through a grant from
the Space Telescope Science Institute, which is operated by the Association of
Universities for Research in Astronomy, Inc., under NASA contract NAS5-26555.

\label{lastpage}
\clearpage


\begin{thebibliography}{}

\bibitem[Alcock et al.(1997)]{alc97}Alcock, C. et al. 1997, ApJ, 486, 697
\bibitem[Alcock et al.(2000a)]{alc00a}Alcock, C. et al. 2000a, ApJ, 541, 734
\bibitem[Alcock et al.(2000b)]{alc00b}Alcock, C. et al. 2000b, ApJ, 542, 281
\bibitem[Aubourg et al.(1993)]{aub93}Aubourg, E. et al.  1993, Nature, 365, 623 
\bibitem[Binney \& Tremaine (1987)]{bin87}Binney, J. \& Tremaine, S. 1987,
           Galactic Dynamics (Princeton: NJ, Princeton University Press)
\bibitem[Blitz \& Spergel(1991)]{bli91}Blitz, L. \& Spergel, D. N. S. 1991, ApJ, 379, 631
\bibitem[Bond et al.(2001)]{bon01}Bond, I. A. et al. 2001, MNRAS, 327, 868 
\bibitem[Bond et al.(2002)]{bon02}Bond, I. A. et al. 2002, MNRAS, 331, L19 
\bibitem[de Vaucouleurs(1964)]{vau64}de Vaucouleurs, G. 1964. IAU Symp. 20. The Galaxy and the Magellanic Clouds, ed. F. J. Kerr \& A. W. Rogers (Canberra: Australian Acad. Science, MSSSO), 195
\bibitem[Eyer \& Wo\'{z}niak (2001)]{eye01}Eyer, L. \& Wo\'{z}niak, P. R. 2001, MNRAS, 327, 601
\bibitem[Eyer (2002)]{eye02}Eyer, L. 2002, Acta Astronomica, 52, 241
\bibitem[Gould(1996)]{gou96}Gould, A. 1996, PASP, 108, 465 
\bibitem[H\"{a}fner et al.(2000)]{haf00}H\"{a}fner, R. et al. 2000, MNRAS, 314, 433
\bibitem[Kiraga \& Paczy\'{n}ski(1994)]{kir94}Kiraga, M., \& Paczy\'{n}ski, B. 1994, ApJ, 430, L101
\bibitem[Mao \& Paczy\'{n}ski(2002)]{mao02}Mao, S. \& Paczy\'{n}ski, B. 2002, preprint (astro-ph/0207131) 
\bibitem[Paczy\'{n}ski(1996)]{pac96}Paczy\'{n}ski, B. 1996, ARA\&A, 34, 419
\bibitem[Paczy\'{n}ski \& Stanek(1998)]{pac98}Paczy\'{n}ski, B. \& Stanek, K. Z. 1998, ApJ, 494, L219
\bibitem[Schechter, Mateo \& Saha(1993)]{sch93}Schechter, L., Mateo, M., \& Saha, A. 1993, PASP, 105, 1342S  
\bibitem[Stanek(1996)]{sta96}Stanek, K. Z. 1996, ApJ, 460, 37L
\bibitem[Stanek et al.(1994)]{sta94}Stanek, K. Z. et al. 1994, ApJ, 429, L73
\bibitem[Stanek et al.(1997)]{sta97}Stanek, K. Z. et al. 1997, ApJ, 477, 163
\bibitem[Stanek et al.(2000)]{sta00}Stanek, K. Z. et al. 2000, Acta Astronomica, 50, 191
\bibitem[Sumi et al.(2002)]{sum02}Sumi, T. et al. 2002, preprint (astro-ph/0207604)
\bibitem[Udalski et al.(1993)]{uda93}Udalski, A. et al. 1993, Acta. Astron., 43, 289
\bibitem[Udalski et al.(1994)]{uda94}Udalski, A. et al. 1994, Acta Astronomica, 44, 165
\bibitem[Udalski et al.(2000)]{uda00}Udalski, A. et al. 2000, Acta Astronomica, 50, 1
\bibitem[Udalski et al.(2002)]{uda02}Udalski, A. et al. 2002, Acta Astronomica, 52, 217 
\bibitem[Udalski, Kubiak \& Szyma\'{n}ski (1997)]{uda97}Udalski, A.,  Kubiak, M., \& Szyma\'{n}ski, M.   1997, Acta Astronomica, 74, 319 
\bibitem[Wo\'{z}niak et al.(2001)]{woz01}Wo\'{z}niak, P. R., et al. 2001, Acta Astronomica, 51, 175
\bibitem[Wo\'{z}niak et al.(2002)]{woz02}Wo\'{z}niak, P. R., et al. 2002, Acta Astronomica, 52, 129
\end{thebibliography}
\end{document}